\documentclass[
tightenlines,
aps,
twocolumn,
floatfix]{revtex4}

\usepackage{graphicx}
\usepackage{float}
\usepackage{amsmath}

\newcommand{\be}{\begin{equation}}
\newcommand{\ee}{\end{equation}}
\newcommand{\ba}{\begin{eqnarray}}
\newcommand{\ea}{\end{eqnarray}}

\newcommand{\ep}{\epsilon}

\begin{document}
\begin{titlepage}

\begin{flushright}
\vbox{
\begin{tabular}{l}
ANL-HEP-PR-11-43
\end{tabular}
}
\end{flushright}
\vspace{0.1cm}

\title{
The four-dimensional helicity scheme and dimensional reconstruction
}

\author{Radja Boughezal}
\email[]{rboughezal@hep.anl.gov}
\affiliation{High Energy Physics Division, Argonne National Laboratory, Argonne, IL 60439, USA}

\author{Kirill Melnikov }
\email[]{melnikov@phys.jhu.edu}
\affiliation{Department of Physics and Astronomy, Johns Hopkins University, Baltimore, MD, USA}

\author{Frank Petriello}
\email[]{f-petriello@northwestern.edu}
\affiliation{Department of Physics \& Astronomy, Northwestern University, Evanston, IL 60208, USA \\
High Energy Physics Division, Argonne National Laboratory, Argonne, IL 60439, USA}

\begin{abstract}
\vspace{2mm}

The four-dimensional helicity  regularization scheme is
often used in  one-loop QCD computations. It was recently argued
in Ref.~\cite{Kilgore:2011ta} that this scheme is inconsistent
beyond the one-loop order in perturbation theory.
In this paper, we clarify the reason  for this inconsistency by studying
the perturbative expansion of the  vector
current correlator in one-flavor QED through three-loop order.
We develop a simple, practical way to fix the four-dimensional
helicity scheme using  the  idea of dimensional reconstruction,
and demonstrate its application in  several illustrative examples.
\end{abstract}

\maketitle

\thispagestyle{empty}
\end{titlepage}

The four-dimensional helicity (FDH) regularization scheme~\cite{bk,Bern:2002zk}
is one of several regularization schemes~\cite{'tHooft:1972fi,collins,siegel}
based on the idea that
consistent definitions of
quantum field theories can be achieved
through analytic continuation in the
number of space-time dimensions~\cite{'tHooft:1972fi}.
The modern use of the FDH scheme is largely
restricted to one-loop computations~\cite{oneloop}.
The motivation for FDH
arose from on-shell methods for loop computations~\cite{Bern:1994zx},
which seek to reconstruct
higher-loop scattering
amplitudes  from tree-amplitudes through their unitarity cuts.
The tree amplitudes in massless QCD
have a remarkably simple form~\cite{Parke:1986gb}
if a four-dimensional
concept -- the spinor-helicity
formalism~\cite{Berends:1981rb}
-- is employed  in  their evaluation.  If this simplification is
to be used  in loop computations, the spin degrees of freedom for
virtual particles must  be treated as four-dimensional, in contrast
to their momenta.  This distinction is made manifest in the FDH scheme.

Until recently, very little work was done to extend the FDH scheme
beyond one-loop computations. To the best of our knowledge,
the majority of complete multi-loop  computations in non-supersymmetric
theories are performed  using conventional dimensional regularization
(CDR)~\footnote{In what follows, we do not distinguish between
the t'Hooft-Veltman regularization scheme and CDR.}. In contrast to this, the known higher-order FDH results include some two-loop
scattering amplitudes (see Ref.~\cite{Bern:2002zk} for examples),
that have not been used for a computation  of any  physical quantity.

There is a good  reason for this state of affairs. The CDR scheme is a
natural scheme to use if inclusive quantities such as cross-sections and
decay rates are computed using the optical theorem.  Since computations
of multi-loop integrals in those cases are mostly based on the
integration-by-parts identities \cite{Tkachov:1981wb,Chetyrkin:1981qh},
it is important  to set up calculations in such a way that
$D$-dimensional Lorentz invariance,
where $D$ is  the space-time dimension of CDR,
is explicit in all stages of the
computation. The fact that intermediate observable states are more naturally
described by four-, rather  than $D$-dimensional quantum fields, is
immaterial within such an approach.  The distinction between observable
and unobservable states is accomplished indirectly, by taking the imaginary
part of an appropriate Green's function at the very end of the calculation.

However, CDR  is also used  in existing fully
differential next-to-next-to-leading order (NNLO)
computations~\cite{Anastasiou:2005qj,Melnikov:2006kv,
GehrmannDeRidder:2007hr,Anastasiou:2005pn,Catani:2007vq,Melnikov:2008qs,Catani:2009sm}.  Its use in such situations is much less natural.  CDR  necessitates calculations of multi-parton matrix elements
to higher orders in $\epsilon=(4-D)/2$.
If  FDH were extended to the two-loop order, this  step could be avoided,
leading to increased efficiency in computations of
quantities with rich multi-parton kinematics. In fact, FDH
is a scheme of choice in many calculations that address next-to-leading
order (NLO) QCD corrections to kinematic distributions at hadron colliders (see Ref.~\cite{Binoth:2010ra} for a recent summary of results).
Having in mind that extension of perturbative computations
for some  basic LHC processes to NNLO
is desirable for several reasons, it is interesting to understand
if the FDH scheme can be used beyond one loop in non-supersymmetric theories.

In a recent work~\cite{Kilgore:2011ta}, Kilgore made a step in this direction
by studying
the application of several regularization schemes
to higher-order calculations: CDR, FDH, and the dimensional reduction approach~\cite{siegel} commonly
used in supersymmetric theories.
He considered the imaginary part of the correlator of
two vector currents and pointed out that the FDH scheme as formulated
in ~Ref.~\cite{Bern:2002zk} becomes inconsistent at higher orders.

The goal of this article is to elucidate the reasons behind this inconsistency, and see if
they can be fixed.  We find that the problem with the
FDH scheme follows from the fact that  the gauge invariance of
the full theory  is broken by the restriction of the loop momenta
to a smaller dimensionality than the spin dimensions.  As the result of
this, the Ward identities are not satisfied for the additional spin degrees of freedom.  The situation with the FDH scheme becomes very similar to that  of  the dimensional reduction.
However, we also find that the ills of FDH at NNLO  can be cured very
simply using ``dimensional reconstruction": if the one-loop result for
an observable is known for an arbitrary number of spin dimensions,
then the (incorrect) two-loop FDH result can be fixed once the one-loop renormalization
constants of the theory are known for two different integer numbers of spin dimensions.
No two-loop  CDR computation  is
necessary, so that this set-up preserves some
of the simplicity of the original FDH.  The spirit of
this fix is similar to a technique employed
to reconstruct the rational parts of one-loop amplitudes in generalized
$D$-dimensional unitarity~\cite{Giele:2008ve}.

We would like to illustrate this idea by considering
as simple a  set-up as possible.  We choose to study
Quantum Electrodynamics (QED) with a single massless
fermion field. The Lagrangian of the theory reads
\be
L = \bar \psi ( i \partial_\mu \gamma^\mu  + e A_\mu \gamma^\mu ) \psi
- \frac{1}{4} F_{\mu \nu} F^{\mu \nu},
\label{eq1}
\ee
where $F_{\mu \nu} = \partial_\mu A_\nu - \partial_\nu A_\mu$.
The vector current in this theory
$J^\mu = \bar \psi \gamma^\mu \psi $
is  conserved, $ \partial^\mu J_\mu = 0$, and does not require renormalization.
We study the  correlator of two vector currents,
\be
\omega_{\mu \nu}(q)\Pi(q^2) = - i
\int {\rm d}^4 x e^{iqx} \langle 0| T J_{\mu}(x) J_\nu(0) |0 \rangle ,
\label{eq2}
\ee
where we used $\omega_{\mu \nu}(q) = (q^2 g_{\mu \nu} - q_\mu q_\nu)$.
It is important to stress that, as it is customary in FDH, the external Lorentz indices $\mu$ and
$\nu$ are taken to be four-dimensional.
Conservation of the vector current  requires  that
the imaginary part of the correlator ${\rm Im}[ \Pi(q^2)]$
is {\it finite without
any renormalization}  if the gauge-invariant subset of diagrams  with closed fermion loops (the only
contribution to the renormalization of the electric charge $e$)
are discarded. It is this absence of any renormalization that makes
this quantity an ideal laboratory for our investigation into various
regularization schemes. We note that Kilgore studied the correlator
of the two conserved vector currents in QCD \cite{Kilgore:2011ta}, where
disentangling the coupling constant renormalization is possible, but more
difficult. We believe that focusing on the QED aspect of the problem allows
us to illustrate the main issue very sharply.

It is straightforward to compute  $\Pi(q^2)$ to three-loops
using various regularization  schemes.  We have done this in the variety
of ways, including utilizing {\sf Mincer} \cite{mincer} or  {\sf Air}
\cite{Anastasiou:2004vj}, as well as using in-house implementations
of the Laporta algorithm~\cite{Laporta:2001dd} established previously~\cite{Boughezal:2011vw}.
We find complete agreement with the
result reported in Ref.~\cite{Kilgore:2011ta} when it is
truncated to QED and all terms that are proportional to the number of lepton flavors
are dropped.   The imaginary parts of the correlator
computed in CDR and FDH are
\be
\begin{split}
& {\rm Im}  \left [ \Pi(q^2) \right ]^{\rm CDR}  =
\frac{1}{12 \pi} \Bigg [
1 + \frac{3}{4}  \left ( \frac{\alpha}{\pi} \right )
-\frac{3}{32}\left ( \frac{\alpha}{\pi} \right )^2
  \Bigg ],
\\
& {\rm Im}  \left [ \Pi(q^2) \right ]^{\rm FDH}  =
\frac{1}{12 \pi} \Bigg [
1 + \frac{3}{4} \left ( \frac{\alpha}{\pi} \right )
-\frac{15}{32}   \left ( \frac{\alpha}{\pi} \right )^2
\Bigg ].
\end{split}
\label{eq0}
\ee
The difference is striking.  It implies that the computation
of a {\it finite quantity, that does not require any renormalization},
leads to different results when two different regularization
schemes are applied. Moreover, we emphasize that both CDR and FDH
computations are consistent  with the conservation of the vector current $J_\mu$ in four dimensions
so there is nothing at this point that makes
either of the two  results  in Eq.~(\ref{eq0}) obviously
incorrect.  One could have suspected that a {\it finite}
shift in the coupling constant
-- familiar  from the application of the FDH scheme in one-loop QCD computations
\cite{oneloop}-- can account for the difference of the two results. However,
the known shift of the coupling constant is purely non-abelian \cite{oneloop}.
Since it   vanishes in the abelian (QED) limit, it is not possible
to reconcile the two results shown in Eq.~(\ref{eq0}) by existing means.

To understand the reason behind  the difference, we review the rules~\cite{Bern:2002zk} that
are used in the FDH computation.  We  begin by
considering QED in a $D_s$-dimensional space, but with all momenta
restricted to a $D$-dimensional subspace of this $D_s$-dimensional space.
This arrangement requires $D_s > D$.
Upon performing spin algebra in all contributing diagrams, we take
$D_s \to 4$,  keeping $D$ fixed.
The limit $D \to 4$ is taken at the end of the calculation. We note that
the CDR scheme can be formulated in a similar way, making   it explicit
that the two schemes differ by  the order of limit-taking. Indeed,
to arrive at the CDR result, we take  $D_s \to D$ for fixed $D$,
 and then take the limit $D \to 4$.

The origin of the differences  in CDR and FDH results
can be best understood by presenting $\Pi(q^2)$
in a form  where $D_s$ is kept fixed, while
the limit $D \to 4-2\ep$  is taken. We find
\be
\begin{split}
& {\rm Im}  \left [ \Pi(q^2) \right ]  =
\frac{1}{12 \pi} \Bigg [
1 + \left ( \frac{3}{4} - \frac{3}{8} \delta_{s} \right )
\left ( \frac{\alpha}{\pi} \right )
\\
& + \left ( -\frac{15}{32} - \frac{3}{16} \frac{\delta_{s}}{\ep}
- \frac{3}{32} \frac{\delta_{s}^2}{\ep} + {\cal O}(\delta_s)
\right )
\left ( \frac{\alpha}{\pi} \right )^2
  \Bigg ],
\end{split}
\label{eq3}
\ee
where $\delta_s = D_s - 4$.  The FDH result is obtained by setting $\delta_s = 0$ in Eq.~(\ref{eq3}), while the CDR result corresponds to setting
$\delta_s = -2\ep$ and taking the limit $\ep \to 0$.  ${\cal O}(\delta_s)$ terms that are not enhanced by inverse powers
of $\epsilon$ are present at NNLO but are irrelevant for both CDR and FDH.   A similar term at NLO is  also irrelevant for both CDR and FDH but, as we will
see, it is important for
understanding differences at NNLO between the two schemes.
For this reason, it is  shown explicitly in Eq.~(\ref{eq3}).

It follows from Eq.~(\ref{eq3}) that
the difference between the CDR and FDH schemes appears at NNLO because terms of the form
$\alpha^2 \delta_s/\ep$ are present in that order of the perturbative expansion.
Those terms  either contribute to the final result (CDR), or are
set to zero by convention (FDH). Note that no $\delta_s/\epsilon$ term appears at
NLO.  Therefore, to understand the difference between CDR and FDH schemes,
we must explain why divergent terms proportional
to the number of ``extra-dimensional'' degrees of freedom appear at NNLO.

The reason becomes very clear if we set $D_s$ to an integer value greater
than four.  For the sake of argument, we take $D_s = 5$.
It immediately follows from Eq.~(\ref{eq3}) that  ${\rm Im} \,\Pi(q^2)$
is  {\it divergent}.   To see why this divergence
occurs, we must go back to the QED Lagrangian in Eq.~(\ref{eq1}) and
ask  what  theory arises if we set $D_s$ to five but keep
all space-time coordinates   four-dimensional.

We begin by extending the Dirac algebra to five dimensions by taking $\Gamma^\mu
= \gamma^\mu$, $\mu = 0,..,3$ and $\Gamma^4 = i \gamma_5$, so that
\be
\Gamma^{M} \Gamma^{N} + \Gamma^{N} \Gamma^{M} = 2 g^{M N},\;\;\;
M,N = 0,..4.
\ee
The fermion fields are not analytically continued, so the number of
independent fermion helicities remains two.  The gauge field
$A^M$ is split into a four-dimensional gauge field and a scalar
field,   $A^M = (A^\mu,\phi)$.
The QED Lagrangian of Eq.~(\ref{eq1}) written in terms of four-dimensional
fields reads
\be
\begin{split}
L = & \bar \psi ( i \gamma_\mu \partial^{\mu} + e A_\mu \gamma^\mu ) \psi
- i g_\phi \bar \psi \gamma_5 \psi \phi
\\
& -\frac{1}{4} F_{\mu \nu} F^{\mu \nu}
+ \frac{1}{2} \partial_\mu \phi \partial^\mu \phi.
\end{split}
\label{eq4}
\ee
Note that in Eq.~(\ref{eq4})
we introduced a new coupling constant $g_\phi$, to parameterize
the interaction  of the field $\phi$ and the pseudoscalar fermion current.
Because the Lagrangian in Eq.~(\ref{eq4}) originates from the five-dimensional
QED Lagrangian in Eq.~(\ref{eq1}), $g_\phi = e$.
However, since we use four-dimensional momenta and coordinates,
this equality of the coupling constants  can  not be protected by the full
$D_s$-dimensional  gauge invariance.  This implies that in Eq.~(\ref{eq4}),
the coupling constant $g_\phi$ requires renormalization, while the electric charge
$e$ is not renormalized and is
protected by the four-dimensional gauge invariance.  For $D_s = 5$,
the result shown in Eq.~(\ref{eq3}) corresponds
to the calculation of ${\rm Im}[\Pi(q^2)]$ in a theory defined
by Eq.~(\ref{eq4})  in terms of {\it bare charges}, $e$ and $g_\phi$.
The bare electric charge $e$ coincides with the physical charge because of the
four-dimensional gauge invariance, but renormalization is required for $g_\phi$ to make
the correlator of the two vector currents explicitly finite in $D_s =5$.  Such renormalization
has not been performed in Eq.~(\ref{eq3}).  This is the reason for the $1/\ep$ divergences present there.
We conclude   that the ``divergences'' in Eq.~(\ref{eq3}) --
crucial for understanding  the CDR/FDH difference --
can be related to the renormalization of the coupling constant
$g_\phi$ for finite $D_s$. Below we describe the details of this relation.

Because the scalar field $\phi$ contributes to the correlator
of the two vector currents only at NLO, through NNLO
we only need the one-loop renormalization
of the coupling constant $g_\phi$. It is easy to obtain this renormalization constant by
considering the Green's function $\langle 0 | T \bar \psi (x) \phi \psi(x) |0
\rangle $. We find
\be
\alpha_\phi^{\rm bare} \left |_{D_s = 5} \right.
 = \alpha \left ( 1 -  \frac{3}{4 \ep } \frac{\alpha}{\pi}
\right),
\label{eq4a}
\ee
where $\alpha_\phi = g_\phi^2/4\pi$ is introduced.
Rewriting Eq.~(\ref{eq3}) for $D_s = 5$ ($\delta_s\to 1$) and separating
the two couplings at NLO explicitly, we find
\be
\begin{split}
& {\rm Im}  \left [ \Pi(q^2) \right ]_{D_s = 5}  =
\frac{1}{12 \pi} \Bigg [
1 + \frac{3}{4} \left ( \frac{\alpha}{\pi} \right )
- \frac{3}{8} \left ( \frac{\alpha_\phi^{\rm bare}}{\pi} \right )
\\
& + \left ( -\frac{15}{32} - \frac{9}{32 \ep} \right )
\left ( \frac{\alpha}{\pi} \right )^2
  \Bigg ].
\end{split}
\label{eq5}
\ee
Removing the bare coupling from Eq.~(\ref{eq5}) using Eq.~(\ref{eq4a}), we
see that the divergence in Eq.~(\ref{eq5}) disappears.  This proves
our assertion about the origin of the divergent
$\delta_s/\ep$ terms in Eq.~(\ref{eq3}).

Having understood the origin of divergences in Eq.~(\ref{eq3}), we must
find a way  to calculate the difference between ${\rm Im}[\Pi(q^2)]$
in the  FDH and CDR schemes without performing a complete three-loop computation.
We observe in Eq.~(\ref{eq3}) that
only the ${\cal O}(\delta_s/\ep)$ term contributes to the CDR/FDH
difference;  the ${\cal O}(\delta_s^2/\ep)$ term is not relevant.   However, since Eq.~(\ref{eq3})
is a second-degree polynomial in $\delta_s$, it is not possible to isolate the desired term  by performing the computation in a single
integer-dimensional space.  Two such calculations are required.  A similar need occurs when attempting to reconstruct the rational parts of
one-loop amplitudes using tree-level amplitudes in higher integer dimensions~\cite{Giele:2008ve},
albeit for a different reason.

We have already discussed the case
$D_s = 5$.  The case $D_s = 6$ is qualitatively similar, but different in detail.
The $D_s=6$ QED Lagrangian of Eq.~(\ref{eq1}) deconstructs to
\be
\begin{split}
L = & \bar \Psi ( i \gamma_\mu \partial^{\mu} + e A_\mu \gamma^\mu ) \Psi
- g_\phi \sqrt{2} \left ( \bar \Psi \gamma_5 \sigma^+ \Psi \phi +
h.c \right )
\\
& -\frac{1}{4} F_{\mu \nu} F^{\mu \nu}
+ \partial_\mu \phi \partial^\mu \phi^*,
\end{split}
\label{eq6}
\ee
where $\bar \Psi = (\bar u, \bar d) $ is the  lepton ``doublet'',
$\sigma^+ = (\sigma^{1}+i\sigma^{2})/2$, $\sigma^{1,2,3}$ are the Pauli
matrices and $\phi$ is a complex scalar field.
We define the conserved vector current as
$J_\mu = ( \bar u \gamma_\mu u + \bar d \gamma_\mu d)/\sqrt{2}$, where
the normalization factor is chosen for convenience.
The corresponding result for the imaginary part of the polarization operator
follows from Eq.~(\ref{eq3}), where we isolate the contribution due to scalar
degrees of freedom at one-loop:
\be
\begin{split}
& {\rm Im}  \left [ \Pi(q^2) \right ]_{D_s=6}  =
\frac{1}{12 \pi} \Bigg [
1 + \frac{3}{4} \left ( \frac{\alpha}{\pi} \right )
- \frac{3}{4} \left ( \frac{\alpha_\phi^{\rm bare}}{\pi} \right )
\\
& + \left ( -\frac{15}{32} - \frac{3}{4 \ep} \right )
\left ( \frac{\alpha}{\pi} \right )^2
  \Bigg ].
\end{split}
\label{eq7}
\ee
The divergence is removed by the renormalization of the bare coupling
$g_\phi$ which, for $D_s = 6$, is computed from the ``flavor-changing''
Green's function
$ \langle 0 | T \bar d \phi u | 0 \rangle $. We find
\be
\alpha_\phi^{\rm bare} \left |_{D_s = 6} \right. =
\alpha \left ( 1 -    \frac{\alpha}{\pi \ep}
\right ).
\label{eq8}
\ee
It is clear from Eq.(\ref{eq7}) that
this renormalization of the coupling constant
makes ${\rm Im} \left [ \Pi(q^2) \right ]$ finite.

Since we understand the structure of ultraviolet divergences for two values
of $D_s$, it is easy to find a relation between the FDH and CDR schemes.
We imagine that a
one-loop computation is performed, and the $D_s$-dependence of the
one-loop result is established.   We assume that  the two-loop FDH result
is also known. The result for general $D_s$ reads
\be
\begin{split}
&  {\rm Im}  \left [ \Pi(q^2) \right ]^{\delta_s}  =
{\rm Im}  \left [ \Pi(q^2) \right ]^{\rm FDH}
+
\\
& \frac{1}{12 \pi}
\Bigg [
- \frac{3}{8} \delta_s \left ( \frac{\alpha_\phi^{\rm bare}}{\pi} \right )
+ \left (
c_1 \frac{\delta_s}{ \ep}
+ c_2 \frac{\delta_s^2}{\ep}
\right )
\left ( \frac{\alpha}{\pi} \right )^2
  \Bigg ].
\end{split}
\label{eq9}
\ee
The CDR result corresponds to setting
$\delta_s \to -2\ep$ in Eq.~(\ref{eq9}) and neglecting
all ${\cal O}(\ep)$ terms. Doing so, we find
\be
  {\rm Im}  \left [ \Pi(q^2) \right ]^{\rm CDR}  =
{\rm Im}  \left [ \Pi(q^2) \right ]^{\rm FDH}
-\frac{c_1}{6\pi} \left ( \frac{\alpha}{\pi} \right )^2.
\label{eq10}
\ee
The connection between the two schemes requires knowledge of
the coefficient $c_1$.  As we discussed earlier, both
  $c_{1}$ and
$c_2$ are related to the renormalization constants of the
couplings of pseudoscalar fields,
that appear as the result of  dimensional
deconstruction, to fermion bi-linears.
Hence, it is a simple matter to find $c_1$.  We require that Eq.~(\ref{eq9}) becomes {\it finite}
for $D_s = 5,6$ if the renormalization of the coupling constant
$g_\phi$ is performed.  Doing so for both values of $D_s$ leads to a system of two equations that can be solved for $c_1$ and $c_2$.  We find
\be
c_1 = \frac{3 \pi}{4 \alpha} \ep \left (
\delta Z_5  - \frac{1}{2} \delta Z_6 \right ).
\label{eq14_a}
\ee
In Eq.~(\ref{eq14_a}),
we have  introduced the one-loop renormalization constants for
the couplings of the scalar fields to fermions in the compactification of
$D_s$-dimensional QED to four-dimensional space-time:
\be
\alpha_\phi^{\rm bare} \left |_{D_s} \right. =
\alpha \left ( 1 + \delta Z_{D_s} \right ).
\label{eq11}
\ee
Explicit expressions for $\delta Z_{D_s}$ for $D_s = 5,6$ follow from
Eqs.~(\ref{eq4a},\ref{eq8}). Using those results,
we find
\be
12 \pi {\rm Im} \left [
\Pi(q^2)^{\rm CDR}  -
\Pi(q^2)^{\rm FDH}  \right ]
= \frac{12}{32} \left (\frac{\alpha}{\pi}  \right )^2 ,
\ee
in agreement with the explicit computations of Eq.~(\ref{eq0}).  As advertised,
we are able to obtain the correct NNLO  result  from the FDH result
without dealing with $D_s$-dimensional
spin degrees of freedom  at NNLO.

There are several possible directions that one can explore at this point, including how this
picture generalizes to more complicated theories (QCD, massive QED, etc.) or more complicated
observables. Except for a few comments, in this paper  we restrict ourselves
to QED but we study  observables that depend on the mass of the lepton. We show that
the procedure  we introduced in the context of the
vector current correlator is general and
remains valid also in those cases.
%We continue by applying dimensional reconstruction to two other quantities
%in QED, to illustrate the generality of the procedure.
We work with one massive fermion flavor
in both examples.

\smallskip
\noindent
$\bullet$  We begin by computing  the mass renormalization constant in FDH in the on-shell scheme,
and ask if
we can relate  it  to the mass renormalization
constant in CDR. The  mass renormalization constant is defined as
\be
m_0 = Z_m m, \label{ref_eq_bare}
\ee
where $m_0$ is the bare fermion mass and $m$ is pole mass of a lepton.  One can easily
read off the ${\overline {\rm MS}}$ mass renormalization constant
from Eq.~(\ref{ref_eq_bare}) because the lepton pole mass is an infra-red finite quantity.
We compute $Z_m$ through two-loop order in QED. We consistently
neglect the contribution of the fermion loops, so that the electric
charge does not need to be renormalized.  The mass
renormalization constant takes the form
\be
Z_m = 1  + a_0 Z_m^{(1)} + a_0^2 Z_m^{(2)},
\label{eq_zm}
\ee
where $a_0 = \alpha/\pi \Gamma(1+\ep)/(4\pi)^{-\ep} m^{-2\ep}$ and
\be
\begin{split}
& Z_m^{(1)} = - \frac{3}{4\ep} - \frac{5}{4}  -
\frac{\delta_s}{8} \left ( \frac{1}{\ep}  + 1 \right )+{\cal O}(\epsilon),
\\
& Z_m^{(2)} =
 \frac{1}{\ep^2} \left ( \frac{9}{32}  + \frac{\delta_s}{16}
        - \frac{\delta_s^2}{128} \right )
       + \frac{1}{\ep}  \left(  \frac{53}{64}
      +\frac{\delta_s}{16}   \right.
\\
&   \left.
- \frac{13 \delta_s^2}{256}
          \right )
   + \frac{219}{128} - \frac{5\pi^2}{16}
        + \frac{\pi^2}{2} \ln 2 - \frac{3}{4}\zeta_3 +{\cal O}(\epsilon).
\label{eqzm1}
\end{split}
\ee
We have written the result in a form where the $\delta_s$-dependent terms are manifest.
The expression Eq.(\ref{eqzm1}) can be translated into the CDR and FDH values for
the on-shell mass renormalization constant  by setting $\delta_s$ to the appropriate
values.  Suppose we have computed Eq.~(\ref{eqzm1}) using the FDH scheme.
Can we obtain the mass anomalous dimension in the CDR scheme without
doing a complete calculation?

The evolution equation for the mass parameter  reads
\be
\mu \frac{ {\rm d} m(\mu) }{{\rm d} \mu }
= m \left ( 2\ep \alpha + \beta(\alpha) \right )
\frac{\partial}{\partial \alpha} \ln Z_m.
\ee
Taking $Z_m^{\rm FDH}$ from Eq.~(\ref{eqzm1}) and setting $\beta(\alpha) = 0$,
we find
\be
\mu \frac{ {\rm d} m(\mu) }{{\rm d} \mu }
= m \gamma(a) = m \left ( 1 + \sum \limits_{i=1}^{\infty} \gamma_i a^i \right)
\ee
which implies
\be
\gamma_1^{\rm FDH} = -\frac{3}{2},\;\;\; \gamma_2^{\rm FDH} = \frac{53}{16}.
\ee
To find the anomalous dimension in the CDR scheme, we write a relation
between the FDH renormalization constant and the renormalization constant
at arbitrary $D_s$
\be
Z_m(\delta_s) - Z_m^{\rm FDH} =
- a\frac{\delta_s}{8 \ep}
+ a^2 \left (
\frac{c_{21} \delta_s}{\ep^2} + \frac{c_{22}\delta_s^2}{\ep^2} \right )
+ ... ,
\ee
where the ellipses stands for other terms that do not affect the anomalous
dimension. To find $Z_m^{\rm CDR}$, we need $c_{21}$, since it
leads to divergent contribution in the limit $\delta_s = -2\ep$.
Repeating what we did for the
photon vacuum polarization,
we must consider the theory at finite $D_s$, so that $c_{21}$ and
$c_{22}$ contribute to the leading two-loop divergence of the fermion
self-energy. Since such divergence is entirely fixed by the lowest-order mass anomalous dimension and the $\beta$-functions for the coupling
constants, we can find an equation for $c_{21}$ and $c_{22}$.  We note
that the $\beta$-functions appear because of the need to
renormalize the scalar-fermion
couplings, as described in Eqs.~(\ref{eq4a},\ref{eq8}).
The relevant condition is that
\be
2\ep \alpha \frac{\partial}{\partial \alpha} \ln Z_m(\delta_s) +
\beta(\alpha_\phi) \frac{\partial}{\partial \alpha_\phi} \ln Z_m(\delta_s)
\ee
is free from $1/\ep$ singularities for any value
of $\delta_s$. In practice, we choose
$D_s = 5$ and $D_s=6$.  The $\beta$-functions follow from
Eqs.~(\ref{eq4a},\ref{eq8}).  We write them here for completeness: $\beta(\alpha_\phi) = -3/4 a^2$ for $D_s = 5$, and $\beta(\alpha_\phi) = -a^2$ for $D_s = 6$.  We finally find
$c_{21} = 1/16$ and $c_{22} = -1/128$,
in agreement with Eq.~(\ref{eqzm1}).
 The mass anomalous
dimensions in the CDR scheme follows immediately.  Finally,
one can imagine that
the difference between  on-shell
$Z_m$ factors in different schemes
Eq.(\ref{eqzm1}) can be understood completely,
by going beyond the $\overline {\rm MS}$
renormalization of the $g_\phi$ coupling constants
as in Eqs.~(\ref{eq4a},\ref{eq8}) and insisting that
the two couplings $g_\phi$  and $e$ are equal to each other,
including the finite renormalization.
We did not pursue this question in this paper but it is an
interesting avenue for further studies.

\smallskip
\noindent
$\bullet$ As the final  example we  compute the two-loop QED corrections  to the electron anomalous magnetic
moment and show that the correct result can be obtained
using the FDH scheme and the procedure outlined above.
We begin by writing the amplitude for the electron scattering off the electromagnetic field as
\be
\begin{split}
& i {\cal M} = -i e\, \bar{u}(p_2) \Gamma u(p_1),
\\
& \Gamma  = \hat \epsilon F_1(q^2) + \frac{i \sigma^{\mu\nu} \epsilon_ \mu q_{\nu}}{2m} F_2(q^2).
\label{eq_def}
\end{split}
\ee
In Eq.~(\ref{eq_def}), $\epsilon_\mu$ is the ``polarization vector'' of the external field,
$\hat \epsilon = \gamma^\mu \epsilon_\mu$, and
$q=p_2-p_1$ is the  momentum transfer from the electron to the field.
The anomalous magnetic moment is given by $a_e = (g-2)/2 = F_2(0)$.  The one-loop result for
arbitrary $\delta_s$ is given by
\be
a_e^{(1)} = \frac{\alpha}{2\pi} \left(1-\frac{\delta_s}{2} \right).
\label{eq_a}
\ee
The two-loop result for $g-2$ requires the on-shell wave-function
and mass renormalization constants for the electron at the one-loop order.
The mass renormalization constant $Z_m$ is given in Eq.~(\ref{eqzm1}).
The wave-function renormalization constant $Z_2$ coincides
with $Z_m$ in QED at this order
in both CDR and FDH schemes.
We find the following results for the two-loop
contribution to the electron anomalous magnetic moment
in the CDR and FDH schemes
\be
\begin{split}
& a_e^{(2),{\rm CDR}} = \left(\frac{\alpha}{\pi}\right)^2 \left\{ -\frac{31}{16}
+\frac{3}{4} \zeta_3 -\frac{\pi^2}{2} {\rm ln} \,2 +\frac{5 \pi^2}{12}\right\}, \\
& a_e^{(2),{\rm FDH}} = \left(\frac{\alpha}{\pi}\right)^2 \left\{ -\frac{35}{16}
+\frac{3}{4} \zeta_3 -\frac{\pi^2}{2} {\rm ln} \,2 +\frac{5 \pi^2}{12}\right\}.
\end{split}
\ee
Our CDR result matches well-known results in the literature~\cite{Barbieri:1972as,Fleischer:1991xp},
when fermion-loop contributions are neglected.  The FDH result is new.
We now illustrate how to
use dimensional reconstruction to obtain the CDR result,
given the $\delta_s$-dependent 1-loop result in Eq.~(\ref{eq_a})
and the 2-loop FDH result.
We proceed as we did for the current correlator by writing the result for arbitrary $\delta_s$ as
\be
\label{g2deltas}
a_e^{\delta_s} = a_e^{{\rm FDH}} -\frac{\delta_s}{4} \left(\frac{\alpha_{\phi}^{bare}}{\pi}\right)  + \left (
c_1 \frac{\delta_s}{ \ep}
+ c_2 \frac{\delta_s^2}{\ep}
\right )
\left ( \frac{\alpha}{\pi} \right )^2.
\ee
The CDR result is obtained by taking $\delta_s = -2\epsilon$:
\be
\label{FDHfix}
a_e^{{\rm CDR}} = a_e^{{\rm FDH}} -2 c_1 \left ( \frac{\alpha}{\pi} \right )^2.
\ee
To obtain $c_1$, we compute Eq.~(\ref{g2deltas}) for $D_s=5,6$ and demand the result be finite after renormalizing $\alpha_{\phi}^{bare}$.  We obtain
\be
c_1 = \frac{\pi}{4 \alpha} \ep \left ( 2\,
\delta Z_5  - \delta Z_6 \right ).
\ee
Inserting this into Eq.~(\ref{FDHfix}), we derive
the correct (CDR) result for $g-2$. Hence, the procedure
that we developed by studying the correlator of two conserved currents appears to be valid
in a more  general context.

Before concluding, we comment on two possible venues for the extension
of this analysis, namely its extension to QCD and to its application to less inclusive observables.
The first comment concerns the well-established procedure for applying the FDH scheme in
one-loop QCD computations. As explained in Ref.~\cite{oneloop}, it is possible to use FDH in one-loop
computations consistently provided that a finite renormalization of the strong coupling constant,
\be
\alpha_s^{\rm FDH} = \alpha_s^{\rm \overline{MS}} \left ( 1 + \frac{C_A}{6} \frac{\alpha_s}{2\pi} \right ),
\label{eq_al}
\ee
is performed. We can easily understand this result
using dimensional reconstruction idea.  Dimensional reconstruction in QCD  leads to
the appearance of color-octet massless
scalars that interact with both fermions and ``four-dimensional'' gluons.  Tree-level computations
involve four-dimensional fields by definition, and therefore all one-loop amplitudes
are proportional to the ``four-dimensional'' version of the strong coupling constant $\alpha_s$.
Massless QCD is made finite by the coupling constant renormalization which, in the case of dimensional
reconstruction, involves the contribution of color-octet scalars.  Because we only need the divergent
contribution of massless color-octet scalar to the renormalization of $\alpha_s$, we can find it
by inspecting the QCD $\beta$-function,
$
\beta_0 = 11/3 C_A - 2/3N_f - C_A/6 N_s
$
and focusing on the contribution of the color-octet scalars (the term
proportional to $N_s$).
As expected, the required shift in the coupling constant in Eq.~(\ref{eq_al})
and the contribution of the color-octet
scalars to QCD $\beta$-functions are appropriately correlated.  By studying the FDH scheme in one-loop QCD in
terms of dimensional reconstruction, it is obvious that finite renormalization of the coupling
constant in Eq.~(\ref{eq_al}) is the only thing needed to perform self-consistent computations
in FDH \footnote{For processes with hadrons in the initial state, parton distribution functions
in the CDR scheme  are typically employed.  Because of that, FDH results are usually translated to CDR at the
end of the calculation. The rules for such a translation are  given
in Refs.~\cite{oneloop,Signer:2008va}.}.

As a second comment,  it is
interesting to ask what the dimensional reconstruction procedure outlined
in this paper implies  for exclusive
computations. For the sake of argument,
consider again the correlator of two vector currents
in QED. Its imaginary part
is directly related to the inclusive decay rate of a vector
boson. But how should a decay rate be
treated if we require a certain number of ``jets'',
borrowing from the QCD terminology?
At NNLO, it is possible to have
four, three and two jets in the final state. The four-jet rate is
finite at this order.  The three-jet rate is only needed through NLO,
and therefore FDH can be used straightforwardly. The two-jet rate
is needed at NNLO, which makes it obvious that this is the place
where corrections to the inclusive rate must be accommodated.
Moreover, since the phase-space for the two-jet configuration  can
be driven  arbitrarily close to the two-parton kinematics
by appropriate adjustments in the jet selection criteria,
the correction to the inclusive cross-section
is the finite renormalization of the Born matrix element. This argument applies to
processes which possess infra-red finite total cross-sections, but it needs further refinements
for  consistent application of the FDH scheme  to exclusive processes at hadron colliders beyond one-loop.

To conclude, in this paper we explored the four-dimensional helicity
scheme at NNLO, following an interesting observation in
Ref.~\cite{Kilgore:2011ta} that it
becomes inconsistent at that order
in perturbation theory.  To avoid the complications of renormalization,  we studied QED corrections to the imaginary
part of the correlator of two conserved currents. We found that
the differences between the FDH and the CDR schemes
are related to the fact that,
upon continuing QED to a space-time of higher dimensionality while
restricting all the loop momenta to lower-dimensional space-times,
$(D_s-4)$ components of the gauge fields turn into scalar fields and
become unprotected by full $D_s$-dimensional
gauge invariance.  As a result, divergences are introduced
that require additional renormalization.  They are removed
in the FDH scheme by simply ignoring these additional degrees of freedom.  In the CDR scheme, terms of the form $(D_s-4)/\epsilon$ give additional
finite contributions.  One can argue for the
correctness of the CDR result over FDH result by simply stating that the
former respects gauge invariance of the theory in $D_s$-dimensions, while the
latter only respects four-dimensional gauge invariance.  Restoring the
$D_s$-dimensional gauge invariance in the FDH scheme is possible using dimensional reconstruction:
if the one-loop result is known for arbitrary $\delta_s$, and the two-loop FDH result is known,
then the two-loop result in the CDR scheme can be obtained by studying the one-loop renormalization
of the theory in $D_s=5,6$.  This gives a simple, practical prescription for maintaining the simplifying
features of FDH while still getting the answer right~\footnote{We note that this prescription will have to be extended
for yet higher orders in perturbation theory, since the ${\rm N}^k$LO result
for a particular observable is a rank-$k$ polynomial in $D_s$.}.  We
demonstrated the applicability of this procedure
by reconstructing the two-loop CDR results from FDH for three examples:
the correlator of two vector currents in QED,
the mass anomalous
dimension of a fermion in QED, and the electron anomalous magnetic moment. Further studies are required
to extend these
ideas to QCD and develop them to a point where computations of fully exclusive
observables through NNLO in the FDH scheme  are possible and transparent.
This remains an interesting problem for the future.
% but
%we anticipate that these ideas will be useful in future cross-section calculations
%at NNLO in perturbation theory.

\noindent
{\bf Acknowledgments}
K.M. gratefully acknowledges useful discussions with Z.~Kunszt.  K.M. would like
to thank the KITP at UCSB for hospitality during the completion of this paper.
This research is supported by the US DOE under contract
DE-AC02-06CH11357, by the NSF under grants PHY-0855365 and PHY05-51164,
and with funds provided by Northwestern University.

\end{document}